\documentclass[doublespacing]{elsart}
\usepackage{amssymb}
\begin{document}
\begin{frontmatter}
\begin{center}
\title{A simple estimate of the electron hopping energy in the Bechgaard salts} 
\end{center}
\author{V.\v Celebonovi\'c}
\ead{vladan@phy.bg.ac.yu}
\address{Inst.of Physics,Pregrevica 118,11080 Beograd\\
Serbia and Montenegro}

\date{\today}
\newpage
\begin{abstract}
Starting from a previously derived theoretical expression for the electrical conductivity of the Bechgaard salts and experimental data on the resistivity of these materials,it is shown how a value of the electron hopping energy can be determined.Values obtained in this way are in good agreement with estimates made by more complicated methods. 
\end{abstract}
\begin{keyword}
Organic metals \sep Resistivity 
\PACS: 71.10.Pm \sep 72.80.Le   
\end{keyword}
\end{frontmatter}
\newpage
\begin{center}
Introduction
\end{center} 
The family of low-dimensional organic conductors now known as the Bechgaard salts has been discovered in 1980 \cite{JER1}. The general chemical formula of these materials is $(TMTSF)_{2}X$,where $(TMTSF)_{2}$ is bis-tetramethyl-tetrase-
lenofulvalene and $X$ designates various anyons.Transport processes in the Bechgaard salts can not be modelled within the standard theory of metals,and a search for a suitable theoretical description has started a short time after their discovery \cite{JER2} (for a recent review see \cite{JER3}).
Basically,the problem is whether or not the Fermi liquid (FL) picture can be used in theoretical work on the Bechgaard salts.On one side,it is well known that the FL breaks down in strictly one dimensional systems.At the same time the Bechaard salts are experimentally known to be quasi one dimensional (Q1D),which can be taken to imply that the FL is applicable to them.An excellent recent review of the alternative to the FL (the so called Luttinger Liquid model) is avaliable in \cite{LL}. 
    
The aim of the present paper is to combine a recent theoretical expression for the electrical conductivity of the Bechgaard salts,derived within the FL \cite{VL},and experimental data on their resistivity.Starting from these data one can by a simple calculation determine the electron hopping energy in the Bechgaard salts. 

Using the so called "memory function" method and a discrete Hubbard model in one dimension,it was shown in \cite{VL} that the resistivity $\rho$ of the Bechgaard salts can be semi-quantitatively approximated by the following expression: 
\newpage
\begin{eqnarray}
\rho\cong\frac{2 N^{4}\pi\chi_{0}(\omega_{0}^{2}-(b t)^2)}{(t U \omega_{p})^2}[\frac{42.4992}{(1+\exp(\beta(-2t-\mu)))^{2}}
+\nonumber\\ \frac{78.2557}{(1+\exp(\beta(-\mu-2t\cos(1))))^{2}}+\frac{bt}{bt+\omega_{0}}(\frac{4.53316}{(1+\exp(\beta(-2t-\mu)))^{2}}+\nonumber\\\frac{24.6448}{(1+\exp(\beta(-\mu-2t\cos(1))))^{2}})]^{-1}
\end{eqnarray} 

In eq.(1) $t$ is the electron hopping energy,$U$ is the on-site repulsion term in the 1D Hubbard model,$\omega_{p}$ is the plasma frequency,$N$ the number of lattice sites,b is a numerical constant ($b=-1.83879$ from \cite{VL}),$\beta$ is the inverse temperature,$\omega_{0}$ the frequency and $\mu$ the chemical potential.

Experiments on Bechgaard salts usually give as their result the temperature dependence of the resistivity  under predefined external conditions such as the external pressure or the magnetic field.In most experiments, the function $\rho(\beta)$ has a point  $\partial\rho/\partial\beta=0$ for some value of $\beta$.(for example \cite{ISH}). The idea of the calculation to be reported in this paper is to determine the function$(1/\rho)\partial\rho/\partial\beta$.It will be shown that using the measured value of $\beta$ for which this function becomes equal to zero,and parameters of the FL description of the Bechgaard salts,one can calculate the value of $t$,the electron hopping energy.  
\begin{center}   
Calculations
\end{center}
Equation (1) has a very general form.Inserting the function $\mu(\beta,t,n)$ as used in [5] into eq.(1),straightforward algebra leads to the following approximate expression for the resistivity of the Bechgaard salts

\begin{eqnarray}
\rho\cong
0.0610608^{2}{\left(1. + e^
        {\beta\left( -2 t - \frac{\left( -1 + n s \right)t^{6}{\beta}^{6}
         \left|t\right|}{1.1029 + 0.1694 t^{2}{\beta}^{2} + 0.0654 t^{4}{\beta}^4}\right)}\right)}^{2}
\nonumber\\
{\left(1+e^{\beta\left(-\left(\frac{\left(-1+ns\right)t^{6}{\beta}^{6}\left|t\right|}{1.1029 + 0.1694 t^{2}{\beta}^{2} + 0.0654t^{4}{\beta}^{4}}\right)- 2 t \cos (1) \right) } \right) }^{2}\nonumber\\ N^{4}{\chi_{0}}
\left(b t + {\omega_{0}}\right)
\left( -\left( b^{2}t^{2} \right)+{\omega_{0}}^2 \right)
\nonumber\\
\left[t^{2}U^{2}\left(1.45707 bt +
2 bte^{\beta\left(-2t-\frac{\left( -1 + ns \right)t^{6}{\beta}^6
\left|t\right|}{1.1029 + 0.1694t^{2}{\beta }^2 + 
0.0654 t^{4}{\beta}^{4}}\right)}+ ...\right)\right]^{-1}
\nonumber\\
\end{eqnarray}

The meaning of different symbols in eq.(2) is explained after eq.(1). Using eq.(2) in its full form,and expanding the result,one gets an expression for $(1/\rho)\partial\rho/\partial\beta$ which has 2212 terms.Such a result is obviously inapplicable.Selecting the first five terms,one arrives at the following approximate result for $(1/\rho)\partial\rho/\partial\beta$:  
\begin{eqnarray}
\frac{1}{\rho}\frac{\partial\rho}{\partial\beta}\cong\frac{-5.82826 b^{5}t^{6}\exp\beta(-2t-\frac{(-1+ns)\left\langle \left\langle 2\right\rangle\right\rangle\left|t\right|}{1.1029+<<1>>+<<1>>})}{(1+\exp\beta(-2t-\frac{<<1>>}{<<1>>}))^2(1+\exp\beta(<<1>>))^2}\nonumber\\
\frac{1}{(bt+\omega_{0})(<<1>>)^2(\omega_{0}^2-(bt)^2)}-<<1>>-\frac{<<1>>}{<<1>>}-\nonumber\\
\frac{20b^5t^6\exp5<<1>>(-2t-\frac{<<1>>}{<<1>>})}{(1+\exp<<1>>)^2(<<1>>)^4<<1>>(\omega_{0}^2-(bt)^2)}
\end{eqnarray}
where the numbers in $<<...>>$ denote the number of omitted terms.

Although eq.(3) in its full form represents  only the sum of the first five terms of the full expression for 
$(1/\rho)\partial\rho/\partial\beta$ it is still too complicated for any kind of a practical application.However,taking the value of the hopping $t$ as a small parameter,and developing eq.(3) in $t$ up to terms of the order $t^7$ one gets the following approximate result:

\begin{eqnarray}
\frac{1}{\rho}\frac{\partial\rho}{\partial\beta}\cong b^5 (\frac{t}{\omega_{0}})^6[1.36967bt+(0.0830714t\beta-0.393216)\omega_{0}] 
\end{eqnarray}

The parameter $\omega_{0}$ can be expressed as $\omega_{0}=t\Lambda$ with $\Lambda>0$ [5].Within the model used in this work $b=-1.83879$ [5].
It can easily be shown that imposing the condition $(1/\rho)\partial\rho/\partial\beta=0$ on eq.(4) gives a simple expression for the hopping $t$.
\newpage
Solving eq.(4) it follows that:
 
\begin{equation}
t\cong\frac{12.0378(2.51854+0.39322\Lambda)}{\beta\Lambda}
\end{equation}
\medskip

\begin{center}
Discussion and conclusions 
\end{center}

Equation (5) gives the value of the hopping $t$ expressed as a function of the inverse temperature $\beta$ on which $\partial\rho/\partial\beta=0$ and the parameter $\Lambda$.Simple as it is,this expression can have useful applications in experimental work on Q1D organic metals.

The zero point of the function $\partial\rho/\partial\beta$ can be determined in each experimental run.Choosing the value of the parameter $\Lambda$ in analogy with values used in high Tc research (as in [5]),one can by a simple calculation determine the value of the transfer integral $t$.For example,$\Lambda=2.8$ and $\beta=100$ ($\beta=100$ corresponds to a temperature T=116 K) gives $t\cong0.16$eV, which is a realistic estimate,in order of magnitude agreement  with values determined by more complicated methods (for example [3]).If, for a given material,$\partial\rho/\partial\beta=0$ at several values of $\beta$ the calculation discussed here can be performed for each of these points,thus giving the temperature dependence of $t$.     
It is an interesting detail that the band filling $n$ does not enter the final expression obtained above for $(1/\rho)\partial\rho/\partial\beta$.Looking from a purely mathematical point of view,$n$ can be involved in eq.(4) by pushing the developement in series in $t$ up to terms of the order of $t^{11}$.This can be interpreted as meaning that within any given experimental run any possible influence of the change of $n$ on the value of $(1/\rho)\partial\rho/\partial\beta$ is an extremely small effect.

Such a conclusion can be expected on general grounds.
\newpage 
Namely,$n$ is mathematically defined as the band filling,while in experimental work it corresponds to doping of the specimen being studied.In experiments,specimen can be doped before starting en experimental run,but not during it.On the other hand,the calculation described here reffers to conditions during an experimental run,which means to fixed $n$. 

The calculation reported here can be applied to experimental runs at various values of the external pressure.One could thus obtain data on the pressure variation of the value of $t$. Repeating experiments with differently doped specimen (that is,with different values of $n$) and keeping all the other conditions fixed could give an idea on the possible variation of $t$ with doping.Further theoretical work along these lines is already starting.   

\begin{center}
Acknnowledgement
\end{center} 
This contribution has been prepared within the research project 1231 financed by the MNTR of Serbia.

\end{document}